\documentstyle[eqsecnum,multicol,aps]{revtex}
\begin{document}
%%%%%%%%START OF TEXT%%%%%%%%%%%%%%
\title {Nonequilibrium phase transition in the Kinetic Ising model:\\
Is transition point the maximum lossy point ?}
\author {Muktish Acharyya$^+$}
\address{\it Institute for Theoretical Physics\\
University of Cologne, 50923 Cologne, Germany}
\date{\today}
\maketitle

\begin{abstract}
The nonequilibrium dynamic phase transition,
in the kinetic Ising model 
in presence of an oscillating magnetic field,
has been studied both by Monte Carlo
simulation (in two dimension)
and by solving the meanfield dynamical equation
of motion for the average magnetization. 
The temperature variations of hysteretic loss (loop area) and the dynamic
correlation have been studied near the
transition point. The transition point has been identified as
the minimum-correlation point. The hysteretic loss becomes maximum above the
transition point.
An 
analytical formulation has been developed to
analyse the simulation results. 
A general relationship among hysteresis loop
area, dynamic order parameter and dynamic correlation has also been 
developed.

\bigskip
\noindent {\bf PACS number(s): 05.50.+q}
\end{abstract}

%\begin{multicols}{2}
\section{Introduction}
The dynamics of magnetization reversal  
in simple ferromagnetic systems
has recently attracted considerable
scientific interest to study the {\it nonequilibrium} responses.
In this regard,
the dynamical responses of the Ising system in presence of an oscillating 
magnetic field have been studied
extensively  
\cite{rkp,dd,puri,tom,lo,ac}. 
The dynamical hysteretic response 
\cite{rkp,dd,puri} and the nonequilibrium
dynamical phase transition \cite{tom,lo,ac,rik,ma1,ma2} 
are two main subjects of 
interest to study
the dynamic responses of the kinetic Ising model in presence of an
oscillating magnetic field.

Tome and Oliviera \cite{tom} first studied the dynamic transition 
by solving the mean field (MF)
dynamic equation of motion (for the average magnetisation) of the kinetic
Ising model in presence of a sinusoidally oscillating magnetic field.
By defining the order parameter as the time averaged
magnetisation over a full cycle of the 
oscillating magnetic field, they showed that 
the order parameter vanishes
depending upon the value of the temperature and the 
amplitude of the oscillating field. 
In the field amplitude and temperature plane they have drawn
a phase boundary separating dynamic ordered 
(nonzero value of order parameter) and disordered (order 
parameter vanishes) phase. They \cite{tom} have also observed 
and located a {\it tricritical point} (TCP),
(separating the nature (discontinuous/continuous) of the transition)
on the phase boundary line. 

Since, this transition exists even in the static (zero frequency) limit
such a transition, observed \cite{tom}
from the solution of mean field dynamical
equation, can not be dynamic in true sense.
This is because, for the field amplitude less than the
coercive field (at temperature less than the 
static ferro-para transition temperature),
the response magnetisation varies periodically
but asymmetrically even in the zero frequency limit; the system remains locked
to one well of the free energy and cannot go to the other one, in the absence
of noise or fluctuation. On the other hand, in presence of thermal fluctuations,
in the static limit, the system can go from one well to another via the 
formation of nucleating droplets. Vanishingly small field is required to push
the system from one to other well. Consequently, the dynamic phase boundary
collapses, in the presence of thermal fluctuations.

To study the true dynamic phase transition (which 
should disappear in the static
limit) one has to consider the effect of thermal fluctuations. In this regard,
Lo and Pelcovits \cite{lo} first attempted to study the dynamic 
nature of this phase transition (incorporating the effect of fluctuation)
in the kinetic Ising model by Monte Carlo (MC) simulation. 
However, they \cite{lo} have 
not reported any precise phase boundary.
Acharyya and Chakrabarti \cite{ac} studied the nonequilibrium dynamic phase
transition in the kinetic Ising model 
in presence of oscillating magnetic field by 
extensive MC simulation. They \cite{ac} have drawn the phase boundary and 
located a
tricritical point (as observed) on the boundary.
It has been also observed \cite{ac}
that this dynamic phase transition is associated with the 
breaking of the symmetry of
the dynamic hysteresis ($m-h$) loop. 
In the dynamically disordered (value of order parameter vanishes)
phase the corresponding hysteresis loop is
symmetric, and loses its symmetry in the ordered phase (giving
nonzero value of dynamic order parameter).
They 
have \cite{ac} also studied the temperature variation of the ac susceptibility
components near the dynamic transition point.  
It has been observed \cite{ac}
that the imaginary or lossy (real) part of the ac susceptibility gives a
peak (dip) near the dynamic transition point (where the dynamic
order parameter vanishes). It was concluded that this is a possible indication
of the thermodynamic nature of this kind of nonequilibrium dynamical phase 
transition.

The statistical distribution of dynamic
order parameter has been studied by Sides et al \cite{rik}. The nature of the
distribution changes (from bimodal to unimodal)
near the dynamic transition point. They have
also observed \cite{rik}
that the fluctuation of the hysteresis loop area grows and becomes considerably
large as one approach the dynamic transition point.

The relaxation behaviour, of the dynamic order parameter,
near the transition point (in the disordered phase), has been studied
\cite{ma1} recently
by MC simulation and solving meanfield dynamic equation. It has been observed 
that the relaxation is Debye type and the relaxation time diverges near the
transition point. The 'specific heat' and the 'susceptibility' also diverge
\cite{ma2}
near the transition point in a similar manner with that of fluctuations of
order parameter and fluctuation of energy respectively. These observations
\cite{ma2} (divergences of fluctuations)
indirectly supports the earlier facts \cite{rik} where the distribution
of the dynamic order parameter becomes wider and
the fluctuation of hysteresis loop area becomes considerably large near the
transition point.

Recently the experimental evidence \cite{jiang} of dynamic transition
has been found.
The dynamical symmetry breaking (associated to the dynamic 
transition) across the transition point of the 
hysteresis loop, has been observed, 
in highly anisotropic (Ising like) and ultrathin Co/Cu(001) 
ferromagnetic films by surface magneto-optic
Kerr effect, as one passes through
the transition point. 
The dynamical symmetry breaking in the hysteresis loops
has also been observed \cite{suen} in ultrathin Fe/W(110) film.
However, the detailed natures
of the dynamic transition and the phase boundary are not yet studied
experimentally. 

In this communication, the dynamic phase transition has been studied in the 
kinetic Ising model in presence of a sinusoidally
oscillating magnetic field by MC simulation and by
solving the mean field dynamical equation of motion for the average
magnetization. The temperature variations of the hysteresis loss (or loop
area), the dynamic correlation and the phase lag are studied near the
dynamic transition point.
The paper has been
organised as follows: in section II simple analytic forms
are given for the loop area, dynamic correlation and dynamic order 
parameter. In section III a general relationship has been developed
among the various dynamical quantities. In section IV the models are
introduced and in section V the numerical results are given.
The paper ends with a summary of the work in section VI.

\section{Analytic forms of
the loop area and the dynamic correlation near
the transition point}

\noindent The form of the oscillating magnetic field is
\begin{equation}
h(t) = h_0 \cos(\omega t).
\end{equation}

\noindent The dynamic order parameter is defined as
\begin{equation}
Q = {\omega \over 2\pi} \oint m(t) dt,
\end{equation}
\noindent which is nothing but the time averaged magnetisation over a
full cycle of the oscillating magnetic field.
The hysteresis loop area is
\begin{equation}
A = -\oint m dh
 = h_0 \omega \oint m(t) \sin(\omega t) dt,
\end{equation}
\noindent which corresponds the energy loss due to the hysteresis.
The Dynamic correlation is defined as
\begin{eqnarray}
C = <m(t)h(t)> - <m(t)><h(t)>, \nonumber
\end{eqnarray}

\noindent where $<..>$ denotes the time average over the full cycle of
the oscillating magnetic field. Since $<h(t)>$ = 0, one can write

\begin{equation}
C = {\omega \over 2\pi} \oint m(t) h(t) dt 
= {{\omega h_0} \over {2\pi}} \oint m(t) \cos(\omega t) dt.
\end{equation}
\noindent The dynamic correlation has another physical interpretation. For the
cooperatively interacting spin system, this is the negative of the time
averaged spin-field interaction energy (per spin) ($<E_f> = -{{\omega} 
\over {2\pi L^2}}\oint 
\sum_i \sigma_i ~h(t) dt$) over a complete
cycle of the oscillating field.

In the dynamically disordered ($Q = 0$)
phase and near the transition point, 
the time series of the magnetisation ($m(t)$) can be approximated
as a square wave with a phase lag $\delta$ with the applied sinusoidal
magnetic field. 

\begin{equation}
m(t)=
\left\{ 
\begin{array}{rcl}
 1 & ~~{\rm for}~~& 0 < t < \tau/4 + \delta/\omega  \\
-  1 &  ~~{\rm for}~~ &\tau/4+\delta/\omega 
< t < 3\tau/4+\delta/\omega \\
  1 & ~~{\rm for}~~&3\tau/4+\delta/\omega < t < 2\pi/\omega,
\end{array}\right.
\label{dis}
\end{equation}

\noindent where $\tau$ is the time period of 
the oscillating field and $\delta$ is the phase
lag between magnetisation $m(t)$ and the magnetic field $h(t)= h_0 \cos (\omega
t)$. The value of the hysteresis loop area can easily be calculated as
\begin{equation}
A = 4 h_0 \sin(\delta).
\end{equation}
\noindent This form of the loop area
was also obtained \cite{ac} from the assumption that it is approximately
equal to four times the product of coercive field and remanent
magnetization
(here the remanent magnetisation equal to unity),
where the coercive field is identified as $h_0 \sin(\delta)$
(the change in field during the phase lag). Considering the 
same form of the magnetisation the dynamic correlation $C$ can also be
calculated exactly as 
\begin{equation}
C = {{2h_0} \over {\pi}} \cos(\delta).
\end{equation}

\noindent From the above forms of $A$ and $C$ it can be written as
\begin{equation}
{{A^2} \over {(4h_0)^2}} + {{C^2} \over {(2h_0/\pi)^2}} = 1.
\end{equation}
\noindent The above relation tells that the loop area 
$A$ and the dynamic correlation
$C$ is elliptically related to each other. It may be noted here, that 
the previously studied ac susceptibility components \cite{ac} obey a circular
relationship ($\chi'^2 + \chi''^2$ = $(m_0/h_0)^2$), where $m_0$ is the
amplitude of the magnetization.

The ordered region ($Q \neq 0$) can be approximated by considering the
following form of the magnetization

\begin{equation}
m(t)=
\left\{ 
\begin{array}{rcl}
 1 & ~~{\rm for}~~& 0 < t < \tau/4 + \delta/\omega  \\
 1-m_r &  ~~{\rm for}~~ &\tau/4+\delta/\omega 
< t < 3\tau/4+\delta/\omega \\
  1 & ~~{\rm for}~~&3\tau/4+\delta/\omega < t < 2\pi/\omega.
\end{array}\right.
\label{ord}
\end{equation}
\noindent In the above simplified approximation, 
it was considered that the magnetisation
can not jump to the other well, however the value of initial magnetisation is
reduced by the amount $m_r$. 
In the real situation it has been observed that this well is not fully
square (as assumed above in the form of $m(t)$), it has a cusp like 
(or parabolic) shape.
For $m_r$ = 2, the above functional form of $m(t)$ will take the form of
\ref{dis} and one can get the disordered ($Q=0$)
phase. Taking the above form of magnetisation the dynamic order parameter
$Q$ can be calculated as $Q = (2 - m_r)/2$. It may be noted that, in this
simplified approximation the dynamic order parameter $Q$ is independent of
phase lag $\delta$, which is not observed in the real situation (phase lag
shows a peak at the transition point). However, this simple picture can 
anticipate the convex (looking from the origin) 
nature \cite{ac} of the dynamic phase boundary. As the
temperature increases $m_r$ increases and it also increases as the field
amplitude increases. In the simplest asumption, one can consider $m_r$ is
proportional to the product of $h_0$ and $T$. Demanding, $m_r = 2$ for
the dynamic transition ($Q = 0$), one can readily obtain $(h_0)_d T_d$ =
constant. This equation tells that
the dynamic phase boundary will be convex. 
The convex  
nature of the phase boundary remains invariant even
if one
assumes that $m_r$ is any increasing function of 
both $T$ and $h_0$ (for example,
power law; $m_r \sim T^x h_0^y$, in this particular case the equation
of the dynamic phase boundary becomes $T_d^x (h_0)_d^y$ = constant, it
is easy to see that this gives the convex shape
of the dynamic phase boundary ). 
However, this very
simple asumption can not describe the entire form of the phase boundary 
accurately, particularly near the end points ($(h_0)_d$ = 0 
and $T_d$ = 0).

\section{General relation among Dynamic order parameter, Hysteresis loop area
and the Dynamic correlation}

From the usual definitions (given in earlier section) 
of $C$ and $A$, one can write
\begin{eqnarray}
{1 \over {{\sqrt {2\pi}}}}
\left({{2\pi C} \over {\omega h_0}} - i{{A} \over {\omega h_0}}\right) = 
{1 \over {{\sqrt {2\pi}}}}
\oint m(t) \exp
({-i \omega t}) dt, \nonumber
\end{eqnarray}
\noindent where $m(\omega) = {1 \over {\sqrt {2\pi}}} 
\oint m(t) \exp({-i \omega t}) dt$.
\noindent
So,
\begin{flushleft}
$C = {{h_0 \omega} \over {\sqrt {2 \pi}}}{\rm Re}\left(m(\omega)\right)$
\end{flushleft}
\noindent and
\begin{flushleft}
$A = -{h_0 \omega {\sqrt {2\pi}}}{\rm Im}\left(m(\omega)\right)$.
\end{flushleft}

\noindent The general (complex) form of $m(\omega ')$ will be
\begin{flushleft}
$m(\omega') = |m(\omega')| \exp (i\phi)$
\end{flushleft}
\begin{flushleft}
$m(\omega') = {1 \over {\sqrt {2\pi}}}
\left({{4\pi^2 C^2} \over {h_o^2 \omega'^2}} + {{A^2} \over {h_0^2
\omega'^2}}\right)^{1/2} \exp i\left[-\tan^{-1} {A \over {2\pi C}}\right]$
\end{flushleft}

\noindent So, $Q$ is related with $A$ and $C$ as follows
\begin{equation}
Q = {1 \over {\tau}}\oint m(t) dt  
={1 \over {\sqrt {2\pi} \tau}}\int d\omega'
\oint m(\omega') \exp(i\omega' t) 
dt 
={1 \over {2\pi \tau}} \int d\omega' \oint 
{\sqrt {\left({{4\pi^2 C^2} \over {h_o^2 \omega'^2}} + {{A^2} \over {h_0^2
\omega'^2}}\right)}} 
 {\huge {\bf e}}^{ \left[i(\omega't - 
\tan^{-1} {A \over {2\pi C}})\right]} dt.
\label{gen}
\end{equation}
\noindent Above equation gives the general relationship among $Q$, $A$ and $C$. 

It has been observed that the steady response $m(t)$, to a sinusoidally 
oscillating magnetic field ($h(t) = h_0 \cos (\omega t)$), is
periodic (with phase lag $\delta$) and has the same
periodicity ($\tau = 2\pi/\omega$) of the field. So, one can
write $m(t)$ in a Fourier series as
\begin{equation}
m(t)= a_0 + \sum_{n=1}^{\infty} a_n \cos(n\omega t) + \sum_{n=1}^{\infty}
 b_n \sin(n\omega t).
\end{equation}
\noindent From the usual definitions of $Q$, $A$ and $C$, it is easy to see
that 
\begin{flushleft}
$a_0 = Q$,
$a_1 = 2C/h_0$ ~~{\rm and}~~
$b_1 = A/(\pi h_0).$
\end{flushleft}
\noindent So, one can write
\begin{equation}
m(t) = Q + {{2C} \over {h_0}} \cos (\omega t) + .....+{{A} \over {\pi h_0}}
\sin(\omega t) + ....  .
\label{mt0}
\end{equation}
\noindent Keeping only the first 
harmonic terms (ignoring higher harmonics) one can
easily express the instantaneous magnetization as
\begin{equation}
m(t) = Q + m_0 \cos(\omega t - \delta)
\label{mt}
\end{equation}
\noindent where the amplitude of magnetization is $m_0 = [(2C/h_0)^2 + 
(A/(\pi h_0))^2]^{1/2}$ and the phase lag is $\delta = \tan^{-1}(A/(2\pi C))$.

\section{The Model and the simulation scheme}

\subsection{The Monte Carlo study}

The local field (at time $t$) at any site $i$,
of a nearest neighbour ferromagnetic Ising
model in the presence of a time varying external magnetic field $h(t)$
with homogeneous and unit interaction energy can be written as
\begin{equation}
h_i(t) = \sum_j \sigma_j(t) + h(t)
\end{equation}

\noindent where $\sigma_i(t) = \pm1$ and $j$ runs over the nearest neighbour
of site $i$. The local field (at site $i$) $h_i(t)$ has an external field
part $h(t)$, which is oscillating sinusoidally in time
\begin{equation}
h(t) = h_0 \sin(2\pi f t),
\end{equation}

\noindent where $h_0$ and $f$ are the amplitude and frequency of the 
oscillating field.

According to heat-bath dynamics, the probability
$p_i(t)$ for the spin $\sigma_i(t)$ will be up at time $t$ 
is given as
\begin{equation}
p_i(t) = {{e^{h_i(t)/K_BT}} \over {e^{h_i(t)/K_BT} + e^{-h_i(t)/K_BT}}},
\end{equation}

\noindent where $K_B$ is the Boltzmann constant which has been taken
equal to unity for simplicity. It may be noted here that the spin-spin
interaction strength $J$ has been taken equal to unity. The temperature
$T$ is measured in the unit of $J/K_B$. Field is measured in the unit of
$J$.
The spin $\sigma_i(t)$ is oriented (at time $t$) as

\begin{equation}
\sigma_i(t+1) = {\rm Sign} [p_i(t) - r_i(t)]
\end{equation}

\noindent where $r_i(t)$ are independent random fractions drawn from the
uniform distribution between 0 and 1.

In the simulation, a square lattice ($L\times L$) is considered under
periodic boundary conditions. The initial condition is all spins are
up (i.e., $\sigma_i(t=0) = 1$, for all $i$). The multispin coding technique
is employed here to store 10 spins in a computer
word consisting of 32 bits. 10 spins are updated simultaneously
(or parallel) by a single command. All words (containing 10 spins) are 
updated sequentially and one full scan over the entire lattice consists
of one Monte Carlo step per spin (MCSS). 
This is the unit of time in the
simulation. The instantaneous magnetisation ($m(t) = (1/L^2) \sum_i \sigma_i
(t)$) is calculated easily. Some transient loops were discarded to have a
stable loop and all the dynamical quantities were calculated from the stable
loop.

This simulation is performed in a SUN workstation cluster
and the computational speed recorded is 7.14 Million updates 
of spins per second.

\subsection{The Meanfield study}

The meanfield dynamical equation of Ising ferromagnet in the presence of
time varying magnetic field is \cite{tom}
\begin{equation}
{dm \over dt} = -m + {\rm tanh} \left( {{m(t) + h(t)} \over T} \right),
\end{equation}

\noindent where the external time varying field $h(t)$ has the 
previously described sinusoidal form. $T$ is the temperature measured
in the unit of $zJ/K_B$ ($z$ is coordination number and $K_B$ is Boltzmann
constant).
This equation has been solved for
$m(t)$ by fourth order Runge-Kutta method by taking the
initial condition  $m(t=0) = 1.0$. The value of the time differential ($dt$)
was taken $10^{-3}$, so that the error is $O(dt^5) \sim 10^{-15}$.
The frequency $\omega$ of the oscillating field is kept fixed ($\omega =
2\pi$) throughout the study. 
Some transient loops were discarded and all
the values of the response are calculated from a stable loop.

\bigskip

\section{Results}

\subsection{The Monte Carlo results}

In the MC simulation, a square lattice of linear size $L = 1000$ is
considered. The frequency $\omega$ of the oscillating field has been kept
fixed ($\omega = 2\pi\times0.01$) throughout the study. 
From the Monte Carlo simulation technique described above
the $m-h$ or hysteresis loops were
obtained. Some ($\sim 600$) initial transient loops were discarded to have
the stable loop. From this one can easily estimate the length of the 
simulation. For the above choice of frequency, 100 MCSS are required to form
a complete loop (or cycle), and 600 such loops were discarded. It has been
checked carefully that the loop gets stabilised (within a reasonably
useful errorbars) for this choice.
The dynamic order parameter
$Q = {{\omega} \over {2\pi}} \oint m(t) dt$ is readily calculated. 
The loop
area $A$ and the dynamic correlation $C$ have been calculated from the usual
definitions. 
The phase lag $\delta$ (between field and magnetization) has been
calculated by taking the difference between the  
positions of minimum of magnetization and
the magnetic field \cite{ac}.
All values of $
Q$, $A$, $\delta$ and $C$ for a particular temperature were obtained by
averaging over 10 different random samples to obtain the smooth variation.
Fig. 1 demonstrates the dynamic transition (with dynamic symmetry breaking)
and the related phemomena 
(e.g., temperature variations of $A$, $\delta$ etc.) at a
glance. For a fixed field amplitude $h_0 = 0.7$ the time variations of
$h(t)$ and $m(t)$ are plotted for various temperatures in the pictures in
left column and the corresponding $m-h$ loops are shown in the right
column. For very low temperature (topmost pictures of Fig. 1), since no
spin flip occurs (within the time period) the magnetization $m(t)$ remains
constant (unity) and consequently the $m-h$ loop is a straight line having
zero loop area. The dynamic order parameter is unity. 
The concept of phase lag (between $m(t)$ and $h(t)$) is not applicable here.
Obviously the dynamic correlation is zero.
After slight increase
of temperature (pictures in the second row) some small number of spin flips
occurs (within the time period). For some time, $m(t)$ decreases from unity
and again it becomes equal to unity. The phase lag is the frequency ($\omega$)
times the time difference between the positions of minimum
of $m(t)$ and $h(t)$. 
The $m-h$ loop encloses a finite but small
area giving $Q$ less than unity. 
The dynamic correlation starts to grow.
As the temperature increases further the 
phase lag ($\delta$)
and the loop area $A$ increases (pictures in the third row) and the
dynamic order parameter $Q$ decreases. In all three cases, described so far,
the asymmetric shapes of the $m-h$ loops are observed due to asymmetric time
variation of the response magnetisation $m(t)$. The dynamic correlation 
decreases. As the temperature is
very close to the dynamic transition 
temperature, (fourth row), where the time
variation of the response magnetisation is almost symmetric giving maximum
values of $\delta$. The $m-h$ loop is symmetric and the
dynamic order parameter $Q$ is almost zero. 
The dynamic correlation $C$ becomes negative and minimal.
As one increases the temperaure
further (last row), the phase lag decreases, and the loop area decreases.
The dynamic correlation starts to grow further. It may be noted here that
the conventional hysteresis or $m-h$ loop is observed in this region of
temperature. As the temperature increases further the dynamic correlation
grows, shows a maxima or peak and then decreases. The loop area monotonically
decreases. 

The dynamical phase transition, via the dynamical symmetry breaking
of the hysteresis loops,
has been observed in highly anisotropic and ultrathin (2D Ising like) 
ferromagnetic films (Co/Cu(001) and Fe/W(110)) \cite{jiang,suen}
by using surface magneto optic Kerr effect study at room 
temperature. In the 
recent experimental study \cite{suen}
in ultrathin Fe/W(110), the dynamical symmetry
breaking of the hysteresis loop was nicely depicted in Fig. 1 of
Ref.\cite{suen}.

The temperature variations of
$Q$, $\delta$, $C$ and $A$ for two different values of field amplitudes
$h_0$ are shown in Fig. 2. In both the cases, 
it has been observed that, near the dynamic 
transition point ($Q = 0$) the phase lag gives a peak and the dynamic
correlation $C$ gives a shallow dip. 
The dynamic correlation $C$ gives a smeared peak much above 
(around $T = 2.6$)
the static
(ferro-para) transition point ($T_c = 2.269..$) (see Fig. 2). 
The hysteresis loop area $A$ shows a peak
above the dynamic transition point.

It is possible to explain these observations from very simple
analytical 
results described above
(section II). The phase lag $\delta$ becomes maximum near the
dynamic transition point. So,according to the analytical
formulation (for $C$ and $A$) for a fixed value of the
field amplitude as the temperature increases the loop area $A
(= 4h_0 \sin \delta$) starts to
increases as the dynamic order parameter $Q$ starts to decrease and
above the dynamic transition point (complete spin reversal) the loop area will
be maximum and after that $A$ will start to decrease.
Similarly, the dynamic correlation $C$ will remain
approximately equal to zero until a considerable amount of spin flip occurs
and $Q$ changes appreciably and then starts to increase. 
Above and near the transition point, where the
phase lag $\delta$ decreases as temperature increases and $C
={{2h_0} \over {\pi}}\cos(\delta)$ should start to increase. Which has been
observed indeed. However, near the transition point it gives a shallow dip,
where the value of the dynamic correlation $C$ in minimum and negative.
The phase lag $\delta$ should be less than or at most equal to $\pi/2$.
The field ($h(t) = h_0\cos(\omega t)$) crosses zero first at the phase value
$\pi/2$ and it becomes minimum (maximum negative) at the value of phase
($\omega t$) equal to $\pi$. The response magnetisation, should change its
sign (cross zero) within this period. This is true for $\omega \to 0$ limit,
however for finite but sufficiently high frequency, this will not happen.
The phase difference more than $\pi/2$ would be observed yielding the
unconventional shapes of $m-h$ or hysteresis loops.
In practice, it was observed that 
some asymmetric shape of the ($m-h$)
loop gives the value of phase lag $\delta$ slightly higher than $\pi/2$.
In this region, $\cos(\delta = \pi/2 + \epsilon)=-\sin(\epsilon)$, which is 
negative and will show a shallow dip (cusp like shape) at the point
where $\delta$ is maximum. 
According, to the analytical
prediction, the loop area $A (=4h_0 \sin(\delta))$ 
should show maximum at the transition point.
However, strictly speaking and in practice it has been observed that the
loop area $A$ becomes peaked above the transition temperature.
Since the loop area is much more strongly dependent on the actual shape
of the magnetisation (which is not a perfect square wave in the
temperature range concerned here).  
As the field amplitude increases the transition
points shift towards the lower temperature. The maximum of $\delta$ also
increases and consequently the dip of $C$ becomes deeper and it remains
negative over wider range of temperature (since $\delta$ remains larger
than $\pi/2$ over wider range). 
It may be noted that the dynamic correlation $C$ becomes zero (in the 
disordered or $Q$ = 0 region) where the phase lag $\delta = \pi/2=1.57080..$.
The dynamic correlation $C$ shows a smeared peak
at quite higher temperature (above the Onsager value), which was misinterpreted
\cite{neda}
as a signature of the {\it stochastic resonance}. In the MF study
(next section),
it was shown that this is also present in the absence of fluctuations
(or stochasticity).

A similar previous study \cite{ac}, 
showed that the ac susceptibility components
would give peak (or dip) near the transition point. In that case, the 
susceptibility components were calculated from the phase lag $\delta$. The
phase lag $\delta$ would show a peak at the transition point. As a consequence
the susceptibility components would show peak (or dip) reflecting 
the behaviour of phase lag $\delta$. However, in this case, the three 
measurements of phase lag $\delta$, loop area $A$ and dynamic correlation $C$
are completely independent, and indicate the transition point separately.

\bigskip

\subsection{The Meanfield results}

By solving the above meanfield equation the $m-h$ or hysteresis loops were
obtained.
The dynamic order parameter
$Q = {{\omega} \over {2\pi}} \oint m(t) dt$ is readily calculated. The loop
area $A$ and the dynamic correlation $C$ have been calculated by using the
above definitions. The phase lag (between field and magnetization) has been
calculated by taking the difference between the minima 
positions of magnetization and
the magnetic field \cite{ac}. Fig. 3 shows the temperature variations of
$Q$, $\delta$, $C$ and $A$ for two different values of field amplitudes
$h_0$. In both the cases, it has been observed that, near the dynamic 
transition point ($Q = 0$) the phase lag gives a peak and the dynamic
correlation $C$ gives a shallow dip. 
The hysteretic loss $A$ gives peak above the transition (dynamic) point.
The dynamic correlation $C$ gives a smeared peak 
 much above (around $T = 1.3$) the static
(ferro-para) transition point ($T_c = 1.0$) (for a closer view see Fig. 4). 

This high temperature peak 
of the dynamic correlation
was misinterpreted as a signature
of {\it stochastic resonance} \cite{neda}. This peak is indeed present in the
case where the fluctuation is absent
(MF case). The appearance of this peak at 
higher temperature can be explained as follows: for much higher temperature
the time variation of instantaneous magnetisation is 
no longer a square wave like and becomes almost sinusoidal with
a phase lage $\delta$. In a very simple view, it can be 
approximated as $m(t)=m_0 \cos(\omega t - \delta)$ (from eqn \ref{mt}; $Q$ = 0
at very high temperature).
The dynamic correlation becomes $C = {{m_0 h_0} \over 2} \cos(\delta)$. Where
$m_0$ is the amplitude of the magnetisation which monotonically decreases as
the temperature increases. The phase lag $\delta$ also monotonically decreases
in the higher temperature. Consequently $\cos(\delta)$ increases and $m_0$
decreases as temperature increases. So, one would obviously expect a peak
at a finite temperature (high enough) where the competition, between fall
of $m_0$ and rise of $\cos(\delta)$ with respect to the
temperature $T$, becomes comparable.
No stochasticity
is involved in it! 
The loop area $A$ also gives a peak above
the transition point. 
Due to the similar reason given in the earlier section the 
dynamic correlation $C$ gives a shallow
dip near the transition point. 

\bigskip

\section{Summary}

The dynamical response of the kinetic Ising model in presence of a sinusoidally
oscillating magnetic field has been studied both by Monte Carlo simulation
(in two dimension)
and by solving the meanfield dynamical equation of motion
for the average magnetization. 

A general relationship among the hysteresis loop area $A$, dynamic order
parameter $Q$ and the dynamic correlation $C$ has been developed 
(eqn. \ref{gen}).
The time series of the magnetization can be decomposed in a Fourior series
and the constant term is identifed as the dynamic order parameter $Q$, the
amplitudes of first harmonic terms are found to be related to the hysteretic
loss (for sine term) and the dynamic correlation (for cosine term) 
(eqn. \ref{mt0}).

The dynamic order
parameter, the loop area and the dynamic correlation have been 
calculated {\it separately} (both from MC and MF study) and
studied as a function of temperature. It
was observed (in both cases) that the dynamic
correlation shows shallow (negative) dip near the transition point. 
The dynamic transition point has been identified as the
the minimum-correlation point. 
The hysteretic loss $A$ becomes maximum above the dynamic transition point.
In this sense, the dynamic transition point is not the maximum lossy point.
It may be noted that
the earlier study \cite{ac} of the ac susceptibility suggests that the
dynamic transition point would be the
maximum-lossy point, since the imaginary part (or lossy-
part) of the ac or complex susceptibility also shows
a peak near the dynamic transition point. However, there is a remarkable
distinction from the present study. In the earlier study \cite {ac}, the
phase lag was calculated from the simulations and the ac susceptibility
components were calculated from the phase lag. So, it is expected that
the temperature variations of the phase lag will be reflected directly in
the temperature variations of ac susceptibility components. But, in the
present study the measurements of phase lag, dynamic correlation and the
loop area are completely independent. 
This behaviour of the dynamic correlation is 
explained from a simple square wave like 
time variation of the instantaneous response
magnetisation. The oversimplified asumption is incapable of explaining
the peak position (above the transition point) of the hysteretic loss $A$.
However, this simple picture can qualitatively describe the nonmonotonic
temperature variations of $A$ and $C$.

The high temperature (above the static critical point $T_c$) peak of the
dynamic correlation was misinterpreted \cite{neda} as
a signature of the {\it stochastic resonance}. This was also discussed and
an analytical form of the dynamic correlation was proposed to show that the
high temperature peak of the dynamic correlation 
is present even in the absence of
fluctuations (or {\it stochasticity}).

Along with the dynamic correlation, the
dynamic transition can be identified by various thermodynamic
quantities like ac susceptibility \cite{ac}, relaxation time \cite{ma1},
specific heat \cite{ma1}, susceptibility \cite{ma2} and the fluctuations
of dynamic order parameter and energy \cite{ma2}. All these quantities
indicate the thermodynamic natures of this kind of nonequilibrium
dynamic phase transition
by showing peak, dip or divergence near the transition point.

The related phenomena of this kind of nonequilibrium dynamic phase transition
in the kinetic Ising model are mostly based on observations and
not yet analysed by using rigorous theoretical
foundations of equilibrium statistical mechanics available so far. 
The experimental evidences \cite{jiang,suen} are still in the primitive stage.
Experimentally, only the dynamic symmetry breaking of the hysteresis loops
is observed near the transition point. However, the details study of the
nature of the transition, the phase boundary and the associated phenomena
(described above) has not yet done experimentally.

\section*{Acknowledgments}

This work is financially supported by {\bf Sonderforschungsbereich 341}. 
The author would like to thank B. K. Chakrabarti and D. Stauffer for
careful reading of this manuscript.

\bigskip
\begin{center} {\bf Figure Captions} \end{center}
\bigskip

\noindent {\bf Fig.1.} A pictorial demonstration of dynamic transition and
associated phemomena. The figures in the left column represents the time
variation of $h(t)$ and $m(t)$ for different temperatures and the corresponding
$m-h$ loop are shown in right column. Temperature increases from top to
bottom. Monte Carlo results for $L = 1000$, $\omega = 2\pi \times 0.01$ and
$h_0 = 0.7$. 

\bigskip

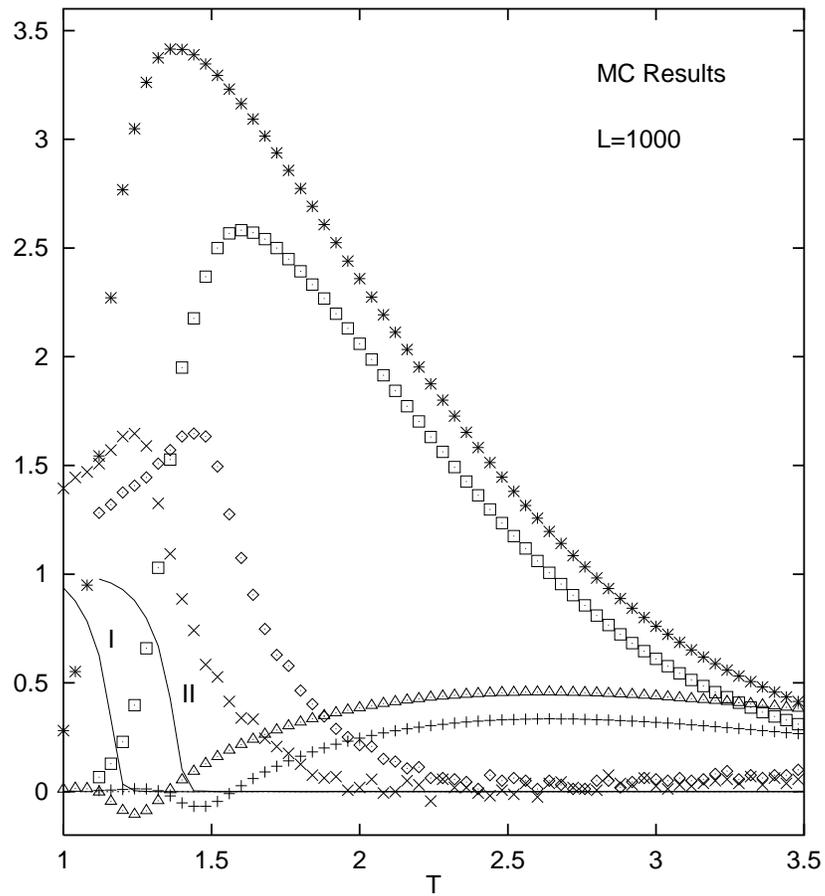
\noindent {\bf Fig.2.} The Monte Carlo results for the temperature 
variations of $Q$, $\delta$, $C$ and $A$ for two different values 
of field amplitudes. $Q$ (solid lines,(I) for $h_0$ = 0.9 and (II)
for $h_0$ = 0.7), $\delta$ ($\times$ for $h_0$ = 0.9 and $\Diamond$ for 
$h_0$ = 0.7), $C$ ($\triangle$ for $h_0$ = 0.9 and $+$ for $h_0$ 
= 0.7) and $A$ ($\star$ for $h_0$ = 0.9 and $\Box$ for $h_0$ = 0.7).

\bigskip

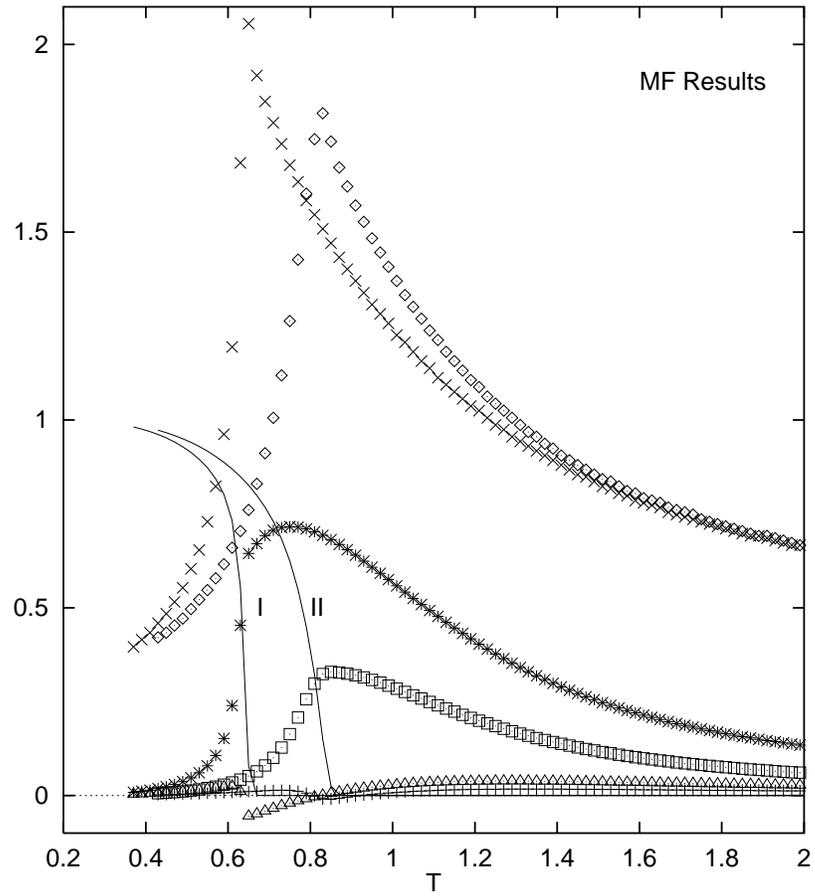
\noindent {\bf Fig.3.} The mean-field results for the temperature 
variations of $Q$, $\delta$, $C$ and $A$ for two different values 
of field amplitudes. $Q$ (solid lines,(I) for $h_0$ = 0.3 and (II)
for $h_0$ = 0.2), $\delta$ ($\times$ for $h_0$ = 0.3 and $\Diamond$ for 
$h_0$ = 0.2), $C$ ($\triangle$ for $h_0$ = 0.3 and $+$ for $h_0$ 
= 0.2) and $A$ ($\star$ for $h_0$ = 0.3 and $\Box$ for $h_0$ = 0.2).

\bigskip

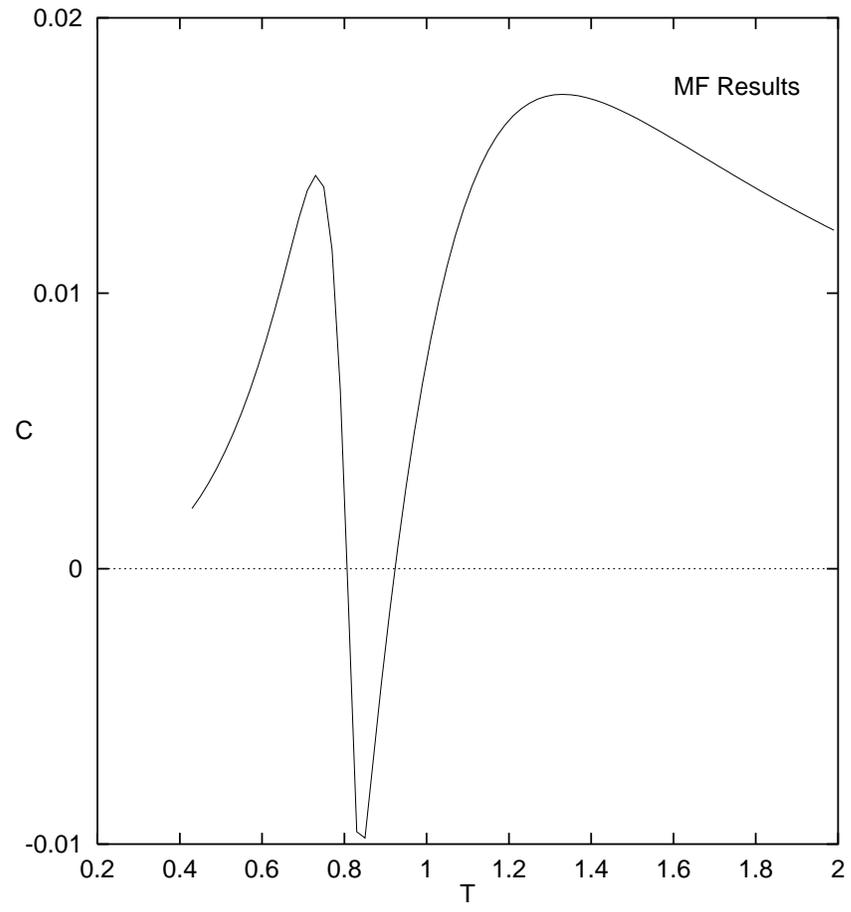
\noindent {\bf Fig.4.} The closer view of Fig.3. for the dynamic 
correlation plotted against the temperature (for fixed field amplitude
$h_0$ = 0.2).

%\end{multicols}

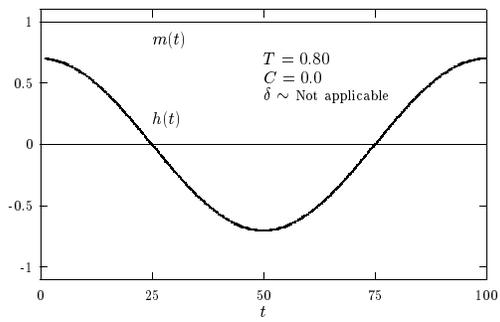
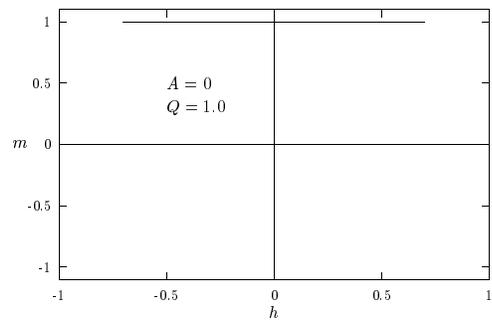
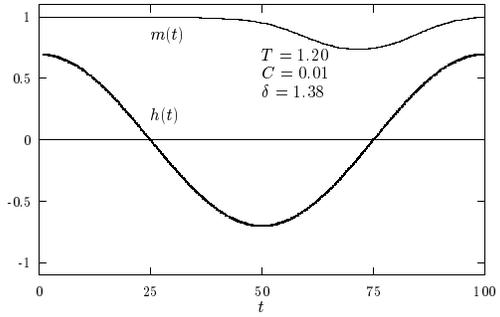
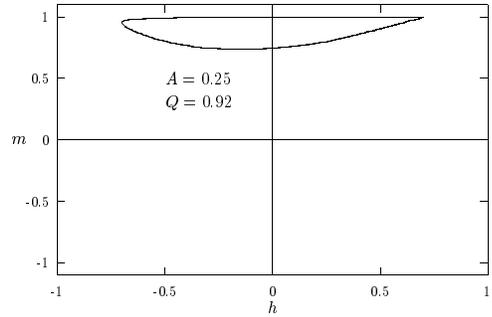
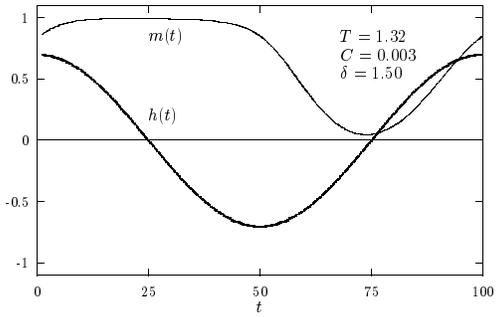
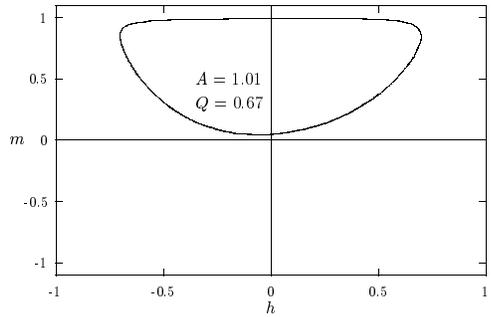
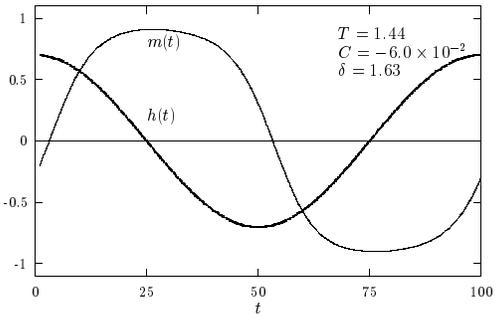
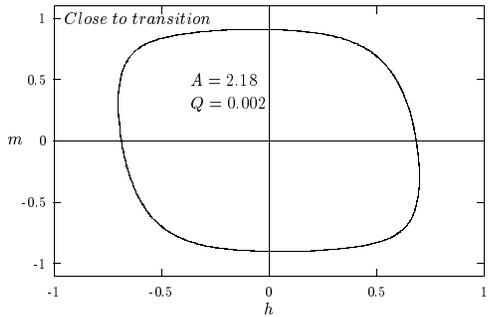
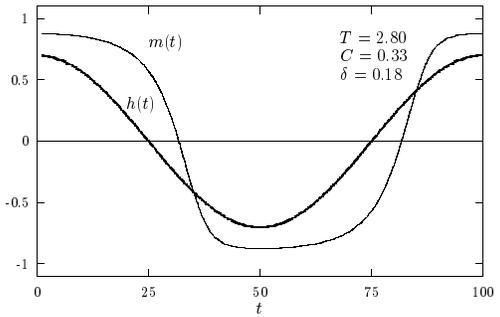
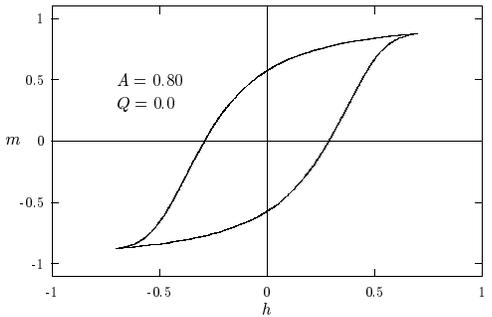
\end{document}